\documentclass[aps,twocolumn,prb]{revtex4}
\usepackage{graphicx}

\begin{document}

\title{Invariant imbedding theory of mode conversion in inhomogeneous
plasmas:
I. Exact calculation of the mode conversion coefficient in cold,
unmagnetized plasmas}

\author{Kihong \surname{Kim}}
\email{khkim@ajou.ac.kr}
\affiliation{Department of Molecular
Science and Technology, Ajou University, Suwon, Korea}
\author{Dong-Hun \surname{Lee}}
\affiliation{Department of Astronomy and Space Science, Kyung Hee
University, Yongin, Korea}

\begin{abstract}
This is the first of a series of papers devoted to the development
of the invariant imbedding theory of mode conversion in
inhomogeneous plasmas. A new version of the invariant imbedding
theory of wave propagation in inhomogeneous media allows one to
solve a wide variety of coupled wave equations exactly and
efficiently, even in the cases where the material parameters change
discontinuously at the boundaries and inside the inhomogeneous
medium. In this paper, the invariant imbedding method is applied to
the mode conversion of the simplest kind, that is the conversion of
$p$-polarized electromagnetic waves into electrostatic modes in
cold, unmagnetized plasmas. The mode conversion coefficient and the
field distribution are calculated exactly for linear and parabolic
plasma density profiles and compared quantitatively with previous
results.
\end{abstract}

\maketitle

\section{Introduction}

The conversion of one type of wave mode into another type of mode at
resonance points in inhomogeneous plasmas and the associated
irreversible transfer of wave energy to the mode conversion region
are important and extensively-studied phenomena in both space and
laboratory plasmas.\cite{budden,ginz,swanson,stix,mjol} There exists
a vast literature on a large number of mode conversion phenomena
involving many different wave modes, which we will not attempt to
review
here.\cite{stix2,moore,chen,woo,cairns,ram,bellan,yin,johnson0,johnson1,
wiles,lee,lee2,denisov,piliya,sc,pert,
means,fors,hink1,hink0,hink2,hink3} Since mode conversion is
associated with the singularity of wave functions, theoretical
studies encounter great difficulties and often adopt some kind of
approximate methods such as the WKB method. In complicated cases
where plasma waves are strongly coupled, these approximate methods
are often unable to produce sufficiently accurate results. Therefore
development of a theoretical method that allows {\it exact}
solutions to mode conversion problems will be of great value in the
study of inhomogeneous plasmas.

In this paper and a series of companion papers to be published
later, we will develop an exact theory of mode conversion based on a
new version of the invariant imbedding method (IIM) for wave
propagation in arbitrarily-inhomogeneous stratified media. The main
idea of the IIM is to transform the original boundary value problem
of wave equations, which are second-order differential equations, to
an initial value problem of coupled first-order ordinary
differential equations for the reflection and transmission
coefficients and the field
amplitudes.\cite{bell,kly0,kly1,rammal,kim1,kim2,kim3} These
equations are called the invariant imbedding equations and the
independent variable, which is usually the thickness of
inhomogeneous media, is called the imbedding parameter. This
transformation makes the numerical solution of wave equations much
easier.

Our IIM utilizes a rigorous integral representation of wave
equations and differs substantially from the better-known IIM
presented in Ref.~29.\cite{lee,lee2,kim1,kim2,kim3} More
specifically, our method gives the invariant imbedding equations
with nonsingular coefficients even in the cases where the material
parameters change discontinuously at some discrete points inside the
inhomogeneous medium, unlike the imbedding equations derived in
Ref.~29. When the inhomogeneity is one-dimensional, which is the
case in stratified media, our method can be used to obtain exact
solutions for the reflection and transmission coefficients and the
electric and magnetic field amplitudes inside
arbitrarily-inhomogeneous media. When the inhomogeneity is random,
it can be used to obtain the exact disorder-averaged reflection and
transmission coefficients and field amplitudes.\cite{kim1} Our
method can also be used to study the propagation of several coupled
waves in both linear and nonlinear media in an exact
manner.\cite{kim3}

In this paper, we apply our IIM to the mode conversion of the
simplest kind, that is the conversion and resonant absorption of
obliquely-incident $p$-polarized electromagnetic waves into
electrostatic modes in cold, unmagnetized
plasmas.\cite{denisov,piliya,sc,pert,means,fors,hink1,hink0,hink2,hink3}
When a $p$ wave of frequency $\omega$ enters a stratified plasma
with a monotonically-increasing density profile at an incident angle
$\theta$, it propagates first to the cutoff point where the local
plasma frequency $\omega_p$ equals $\omega\cos\theta$ and is
partially reflected. Some fraction of the wave then tunnels to the
mode conversion point where $\omega_p$ is equal to $\omega$ and is
converted to electrostatic modes. Since the generation of
electrostatic modes requires the electric field component in the
direction of inhomogeneity, only obliquely-incident $p$ waves can
produce mode conversion in unmagnetized plasmas.

Mode conversion in unmagnetized plasmas has been studied extensively
over many decades and a detailed review of the literature can be
found in Ref.~25. In this paper, we restrict our attention to cold
plasmas with zero temperature and postpone the discussion of finite
temperature effects to future publications. Though our method can be
applied to stratified plasmas with arbitrary density profiles very
easily, we consider only the cases with linear and parabolic density
profiles in this paper. One of the main reasons why we revisit this
well-studied problem is to establish the accuracy and efficiency of
our method before applying it to more complicated mode conversion
problems.

Among a large number of references on the mode conversion in
unmagnetized plasmas with a linear density profile in half- or
entire space, we mention just two papers. The readers may refer to
Ref.~25 for a more complete list of references. Forslund {\it et
al.} performed a numerical simulation of warm, unmagnetized plasmas
at low temperatures.\cite{fors} The mode conversion coefficient and
the field distribution at the lowest temperature they have studied
are generally considered to be an accurate solution of the mode
conversion problem in cold, unmagnetized plasmas.

Hinkel-Lipsker {\it et al.} derived analytical formulas for the mode
conversion coefficient in the cases where the density profile is
linear and parabolic in entire space.\cite{hink1,hink0,hink2,hink3}
Their result in the linear case agrees pretty well with Forslund
{\it et al.}'s result, but there is some numerical
discrepancy.\cite{hink1} In the parabolic case, they considered the
case where the wave frequency is fixed at the value of the local
plasma frequency at the peak of the parabolic density profile
separately from more general cases where the wave frequency is lower
than the peak plasma frequency.\cite{hink0,hink2,hink3}

We calculate the mode conversion coefficient and the magnetic field
distribution for precisely the same models as those considered in
Refs.~24$-$28. We find a very good agreement with the result of
Ref.~24 in the linear case, though we believe that the numerical
accuracy of our result is better. In the parabolic case, we find
substantial differences between our results and those of
Refs.~26$-$28. To the best of our knowledge, the results presented
in this paper represent the first exact solution of the mode
conversion problem in cold, unmagnetized plasmas with linear and
parabolic density profiles.

In Sec.~\ref{sec:model}, we introduce the wave equation and the
plasma density profiles. In Sec.~\ref{sec:imbed}, we present the IIM
and the invariant imbedding equations for the reflection and
transmission coefficients and the field amplitudes. In
Secs.~\ref{sec:lin1} and \ref{sec:lin2}, the results on the case
with a linear density profile in entire or half- space are
presented. In Sec.~\ref{sec:para}, the results on the case with a
parabolic density profile in entire space are presented. We conclude
the paper in Sec.~\ref{sec:conc}.

\section{Model}
\label{sec:model}

We consider a plane, monochromatic and linearly-polarized
electromagnetic wave of frequency $\omega$ and vacuum wave number
$k_0=\omega/c$, incident from a homogeneous region on a stratified
inhomogeneous plasma, where the electron density $n$ and the
dielectric permittivity $\epsilon$ varies only in one direction in
space. We take this direction as the $z$ axis and assume that the
inhomogeneous plasma lies in $0 \le z \le L$ and the wave propagates
in the $xz$ plane. Since the plasma is uniform in the $x$ direction,
the $x$ component of the wave vector, $q$, is a constant and the
dependence on $x$ is taken as being through a factor $e^{iqx}$. When
$q$ is nonzero, the wave is said to pass through the plasma
obliquely.

For $q\ne 0$, we need to distinguish two independent cases of
polarization. In the first case, the electric field vector is
perpendicular to the $xz$ plane and the magnetic field vector lies
in that plane. This type of wave is known as an $s$ (or TE) wave. In
the second case, the magnetic field vector is perpendicular to the
$xz$ plane and the electric field vector lies in that plane. Then
the complex amplitude of the magnetic field, $B=B(z)$, satisfies
\begin{equation}
{{{d^2}
B}\over{dz^2}}-{1\over\epsilon(z)}{{d\epsilon}\over{dz}}{{dB}\over{dz}}
+\left[{{k_0}^2}\epsilon(z)-q^2\right]B=0. \label{eq:p}
\end{equation}
This type of wave is called $p$ (or TM) wave. In Eq.~(\ref{eq:p}),
we have assumed that the magnetic permeability of the plasma is
equal to 1. It is well-known that if the incident wave is
$s$-polarized, mode conversion does not occur in unmagnetized
plasmas. In this paper, we will consider only the $p$ wave case.

It is straightforward to derive an analytical expression
for $\epsilon$ in a cold, unmagnetized electron plasma:
\begin{eqnarray}
\epsilon(z)=1-\frac{{\omega_p}^2}{\omega\left(\omega+i\gamma\right)},~~
{\omega_p}^2=\frac{4\pi e^2}{m}n(z),
\end{eqnarray}
where $\gamma$ is the collision frequency and $\omega_p$ is the
electron plasma frequency. $e$ and $m$ are the electron charge and
mass, respectively. We assume that the wave is incident from region
I where $z>L$ and transmitted to region II where $z<0$. The
dielectric permittivity is equal to $\epsilon_1$ in region I and
$\epsilon_2$ in region II. The phenomenological collision frequency
$\gamma$ is assumed to be zero in regions I and II.

Our method can be applied to the general case where the electron
density $n$ is an arbitrary function of $z$. In this paper, we
restrict our consideration to three special cases. In Model A, the
density depends on $z$ {\it linearly} in {\it entire} space. In
order to treat this case, we assume that the density $n(z)$ is given
by
\begin{equation}
n_A(z)=n_0\left(1-\frac{z-\frac{L}{2}}{\Lambda}\right)
\end{equation}
for $0\le z\le L$ and take the $L\rightarrow\infty$ limit
numerically. The constant $\Lambda$ is the scale length for the
linear density profile. We point out that for $z>\Lambda+L/2$, the
density takes unphysical negative values. When $L\rightarrow\infty$,
$n_A(z)$ varies from $-\infty$ at $z=L$ to $+\infty$ at $z=0$. We
make a further simplification by assuming that the frequency is
fixed at the value of the local plasma frequency at $z=L/2$, which
is $\left(4\pi n_0 e^2/m\right)^{1/2}$, and $\gamma\ll\omega$. We
make these {\it inessential} assumptions just to make our model look
exactly the same as the corresponding model in Ref.~25. Then the
dielectric permittivity takes the form
\begin{equation}
\epsilon_A(z)=\left\{\begin{array}{ll}
\frac{1}{2}\frac{L}{\Lambda} & ~~\mbox{if $z>L$}\\
\frac{z-\frac{L}{2}}{\Lambda}+i\eta
& ~~\mbox{if $0 \le z \le L$}\\
-\frac{1}{2}\frac{L}{\Lambda} & ~~\mbox{if $z<0$}\end{array}\right.,
\end{equation}
where the dimensionless damping parameter $\eta$ is proportional to
$\gamma$ and will be sent to zero numerically. We note that in the
absence of damping, the dielectric permittivity vanishes at $z=L/2$,
where the mode conversion takes place.

In Model B, we assume that the wave is incident from a vacuum on an
inhomogeneous plasma, where the electron density depends on $z$ {\it
linearly} in {\it half}-space. The density is given by
\begin{equation}
n_B(z)=n_0\frac{L-z}{\Lambda}
\end{equation}
for $0\le z\le L$. In the $L\rightarrow\infty$ limit, $n_B(z)$
varies from $0$ at $z=L$ to $+\infty$ at $z=0$. Similarly to the
Model A case, we make an assumption that the frequency is fixed at
the value of the local plasma frequency at $z=L-\Lambda$,
$\left(4\pi n_0 e^2/m\right)^{1/2}$, and $\gamma\ll\omega$. Then the
dielectric permittivity has the form
\begin{equation}
\epsilon_B(z)=\left\{\begin{array}{ll}
1 & ~~\mbox{if $z>L$}\\
1+\frac{z-L}{\Lambda}+i\eta
& ~~\mbox{if $0 \le z \le L$}\\
1-\frac{L}{\Lambda} & ~~\mbox{if $z<0$}\end{array}\right..
\end{equation}
The mode conversion occurs at $z=L-\Lambda$ in this case.

In Model C, we assume that the density depends on $z$ {\it
parabolically} in {\it entire} space and is given by
\begin{equation}
n_C(z)=n_0\left[1-\left(\frac{z-\frac{L}{2}}{\Lambda}\right)^2\right]
\end{equation}
for $0\le z\le L$ and the $L\rightarrow\infty$ limit will be taken
numerically. $n_0$ is the electron density at the top of the
parabolic density profile and $\Lambda$ is the scale length.
Assuming that $\gamma\ll\omega$, the dielectric permittivity is
given by
\begin{equation}
\epsilon_C(z)=\left\{\begin{array}{ll}
\left(\frac{1}{2}\frac{L}{\Lambda}\right)^2 & ~~\mbox{if $z>L$}\\
\left(\delta+1\right)\left(\frac{z-\frac{L}{2}}{\Lambda}\right)^2-\delta+i\eta
& ~~\mbox{if $0 \le z \le L$}\\
\left(\frac{1}{2}\frac{L}{\Lambda}\right)^2 & ~~\mbox{if
$z<0$}\end{array}\right.,
\end{equation}
where $\delta=4\pi n_0e^2/(m\omega^2)-1$. If the frequency is fixed
at the value of the plasma frequency at the top of the density
profile, $\delta$ is equal to zero. When the wave frequency is
shifted downward from the peak plasma frequency, $\delta$ takes a
positive value. Then the mode conversion occurs at
$z=L/2\pm\left[\delta/\left(\delta+1\right)\right]^{1/2}\Lambda$.

We point out that strictly speaking, Models A and C make sense only
in the limit where the damping parameter $\eta$ goes to zero. For
any finite value of $\eta$, the incoming wave of finite amplitude
will decay away before reaching the mode conversion point. When we
present our numerical result for $\eta$ values which are not
sufficiently small, we will always specify the thickness of the
plasma, $L$, used in the calculation.

\section{Invariant imbedding equations}
\label{sec:imbed}

We consider a plane $p$ wave of unit magnitude $\tilde
H(x,z)=H(z)e^{iqx}=e^{ip(L-z)+iqx}$, where
$p=\sqrt{\epsilon_1}~k_0\cos\theta$ and
$q=\sqrt{\epsilon_1}~k_0\sin\theta$, incident on the plasma from the
right (that is, the region where $z>L$). $\theta$ is called the
angle of incidence. The quantities of main interest are the complex
reflection and transmission coefficients, $r=r(L)$ and $t=t(L)$,
defined by the wave functions outside the medium:
\begin{widetext}
\begin{eqnarray}
\tilde H(x,z)&=&\left\{ \begin{array}{ll}
e^{ip(L-z)+iqx}+r(L)e^{ip(z-L)+iqx},  &  ~z>L \\
t(L)e^{-ip^\prime z+iqx},  &  ~z<0  \end{array} \right.,
\end{eqnarray}
\end{widetext}
where $p^\prime$ ($=\sqrt{\epsilon_2}~k\cos\theta^\prime$) is a
positive real constant defined by the dielectric permittivity in
region II, $\epsilon_2$ ($>0$), and the angle that outgoing waves
make with the negative $z$-axis, $\theta^\prime$. If $\epsilon_2<0$,
the wave is not transmitted to region II and the transmission
coefficient needs not to be considered.

Using the IIM, we derive {\it exact} differential equations
satisfied by $r$ and $t$:\cite{lee,kim2,kim3}
\begin{widetext}
\begin{eqnarray}
{{dr(l)}\over{dl}}&=&2i\sqrt{\epsilon_1}~k_0\cos\theta~
\frac{\epsilon(l)}{\epsilon_1}r(l)-{i\over
2}\sqrt{\epsilon_1}~k_0\cos\theta~
\left[\frac{\epsilon(l)}{\epsilon_1}
-1\right]\left[1-\frac{\epsilon_1}{\epsilon(l)}\tan^2\theta
\right]\left[1+r(l)\right]^2,\nonumber\\
{{dt(l)}\over{dl}}&=&i\sqrt{\epsilon_1}~k_0\cos\theta~
\frac{\epsilon(l)}{\epsilon_1}t(l)-{i\over
2}\sqrt{\epsilon_1}~k_0\cos\theta~
\left[\frac{\epsilon(l)}{\epsilon_1}
-1\right]\left[1-\frac{\epsilon_1}{\epsilon(l)}\tan^2\theta
\right]\left[1+r(l)\right]t(l). \label{eq:rtp0}
\end{eqnarray}
\end{widetext}
These equations are supplemented with the initial conditions for $r$
and $t$, which are obtained using the well-known Fresnel formulas:
\begin{eqnarray}
r(0)&=&\frac{\epsilon_2\sqrt{\epsilon_1}\cos\theta
-\epsilon_1\sqrt{\epsilon_2-\epsilon_1\sin^2\theta}}
{\epsilon_2\sqrt{\epsilon_1}\cos\theta
+\epsilon_1\sqrt{\epsilon_2-\epsilon_1\sin^2\theta}},\nonumber\\
t(0)&=&\frac{2\epsilon_2\sqrt{\epsilon_1}\cos\theta}
{\epsilon_2\sqrt{\epsilon_1}\cos\theta
+\epsilon_1\sqrt{\epsilon_2-\epsilon_1\sin^2\theta}}.
\label{eq:icp}
\end{eqnarray}
When $(\epsilon_2-\epsilon_1\sin^2\theta)$ is negative, the square
roots appearing in Eq.~(\ref{eq:icp}) have to be replaced by
$i\sqrt{\vert\epsilon_2-\epsilon_1\sin^2\theta\vert}$.

For given values of $\epsilon_1$, $\epsilon_2$, $k_0$ (or $\omega$)
and $\theta$ and for an arbitrary function $\epsilon(l)$, we
integrate the nonlinear ordinary differential equations
(\ref{eq:rtp0}) from $l=0$ to $l=L$ numerically, using the initial
conditions (\ref{eq:icp}). The reflectivity $\mathcal{R}$ and the
transmissivity $\mathcal{T}$ are obtained by
\begin{eqnarray}
{\mathcal R}=\vert r\vert^2, ~~~ {\mathcal T}=
\frac{\epsilon_1\sqrt{\epsilon_2-\epsilon_1\sin^2\theta}}
{\epsilon_2\sqrt{\epsilon_1}\cos\theta}\vert t\vert^2.
\end{eqnarray}
In Models A and B, we do not need the equations for $t$ and
$\mathcal T$, since the transmission is identically zero in those
cases. In Model C, we note that the dielectric constants in regions
I and II are identical. In that case, the initial conditions are
simplified to $r(0)=0$ and $t(0)=1$ and the transmissivity is given
by ${\mathcal T}=\vert t\vert^2$.

If mode conversion occurs, the energy of the incident wave is
absorbed into the inhomogeneous plasma, even when the damping
parameter vanishes. In Models A and B, the mode conversion
coefficient $\mathcal A$, which measures the wave absorption, is
obtained by ${\mathcal A}=1-{\mathcal R}$. In Model C, $\mathcal A$
is equal to $1-{\mathcal R}-{\mathcal T}$.

The IIM can also be used in calculating the field amplitude $B(z)$
inside the inhomogeneous plasma. We consider the $B$ field as a
function of both $z$ and $L$: $B=B(z,L)$. Then we obtain
\begin{widetext}
\begin{eqnarray}
\frac{\partial B(z,l)}{\partial l}=i\sqrt{\epsilon_1}~k_0\cos\theta
\left\{ \frac{\epsilon(l)}{\epsilon_1}
-\frac{1}{2}\left[\frac{\epsilon(l)}{\epsilon_1}
-1\right]\left[1-\frac{\epsilon_1}{\epsilon(l)}\tan^2\theta\right]
\left[1+r(l)\right]\right\} B(z,l) \label{eq:fd}
\end{eqnarray}
\end{widetext}
For a given $z$ ($0<z<L$), the field amplitude $B(z,L)$ is obtained
by integrating this equation from $l=z$ to $l=L$ using the initial
condition $B(z,z)=1+r(z)$.

\section{Results on Model A: Linear density profile in entire space}
\label{sec:lin1}

The mode conversion in a warm, unmagnetized plasma with a linear
density profile in entire space has been studied analytically by
Hinkel-Lipsker {\it et al.}.\cite{hink1} If we restrict our
attention to the cold plasma case with zero temperature, our Model A
is identical to the model studied in Ref.~25. We introduce the
dimensionless variables
\begin{eqnarray}
{\tilde z}\equiv z/\Lambda,~~{\tilde L}\equiv L/\Lambda,~~{\tilde
l}\equiv l/\Lambda,~~\zeta=k_0\Lambda,
\end{eqnarray}
and solve the invariant imbedding equation for $r$ in
Eq.~(\ref{eq:rtp0}) with the initial condition
\begin{eqnarray}
r(0)&=&\frac{\cos\theta+i\sqrt{1+\sin^2\theta}}{\cos\theta-i\sqrt{1+\sin^2\theta}},
\end{eqnarray}
where we have used the condition $\epsilon_2=-\epsilon_1$.

In the limit where ${\tilde L}\rightarrow\infty$ and
$\eta\rightarrow 0$, we find numerically that the mode conversion
coefficient ${\mathcal A}$ ($=1-\vert r\vert^2$) is a {\it
universal} function of the parameter $Q_a\equiv
\left(k_0\Lambda\right)^{2/3}\sin^2\theta$. That this has to be the
case can be easily seen by rewriting our wave equation using the
variables introduced in Ref.~25, ${\hat z}\equiv {k_0}^{2/3}
z/\Lambda^{1/3}$, ${\hat L}\equiv {k_0}^{2/3}L/\Lambda^{1/3}$,
${\hat\eta}\equiv (k_0\Lambda)^{2/3}\eta$:
\begin{equation}
{{{d^2} B}\over{d{\hat z}^2}}-\frac{1}{{\hat z}-{\hat
L}/2+i{\hat\eta}}{{dB}\over{d{\hat z}}} +\left({\hat z}-{\hat
L}/2+i{\hat\eta}-Q_a\right)B=0. \label{eq:pa}
\end{equation}

\begin{figure}
\includegraphics[width=.4\textwidth]{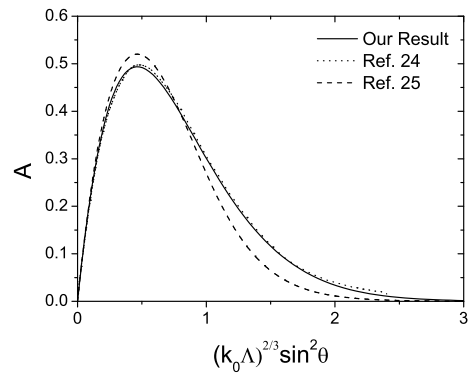}
\caption{Mode conversion coefficient $\mathcal A$ as a function of
$Q_a=\left(k_0\Lambda\right)^{2/3}\sin^2\theta$, for Model A. Our
exact result is compared with the numerical result of Ref.~24 and
the analytical result of Ref.~25.}
\end{figure}

In Fig.~1, we show the mode conversion coefficient as a function of
the variable $Q_a$. In obtaining this result, we have used a
sufficiently large $\tilde L$ ($\sim 100$) and a sufficiently small
$\eta$ ($\sim 10^{-8}$) and the convergence was perfect. We compare
our result with the numerical result of Ref.~24 and the analytical
result of Ref.~25. We find that the agreement with Ref.~24 is very
good, though we believe that the numerical accuracy of our result is
better.

\section{Results on Model B: Linear density profile in half-space}
\label{sec:lin2}

\begin{figure}
\includegraphics[width=.4\textwidth]{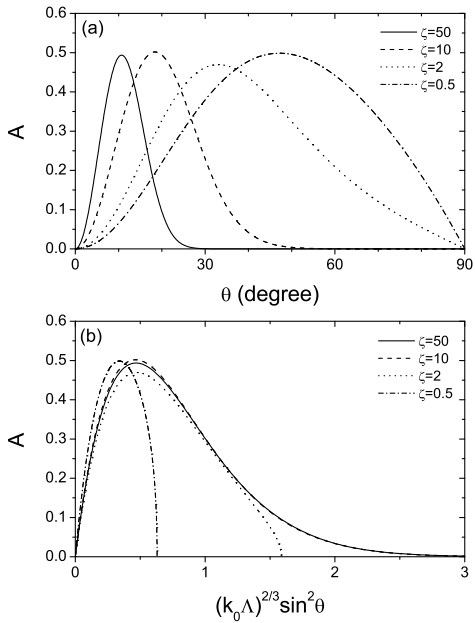}
\caption{(a) Mode conversion coefficient $\mathcal A$ for Model B as
a function of the incident angle $\theta$ for several values of
$\zeta=k_0\Lambda$. (b) $\mathcal A$ as a function of
$Q_a=\left(k_0\Lambda\right)^{2/3}\sin^2\theta$ for Model B. As
$\zeta$ increases, the curves converge to the universal shape in
Fig.~1.}
\end{figure}

In order to obtain the mode conversion coefficient in Model B, we
solve the invariant imbedding equation for $r$ in
Eq.~(\ref{eq:rtp0}) with the initial condition
\begin{eqnarray}
r(0)&=&\frac{\left({\tilde L}-1\right)\cos\theta+i\sqrt{{\tilde
L}-\cos^2\theta}}{\left({\tilde L}-1\right)\cos\theta-i\sqrt{{\tilde
L}-\cos^2\theta}},
\end{eqnarray}
where we have used the values $\epsilon_1=1$ and
$\epsilon_2=1-{\tilde L}$. In this case, the mode conversion
coefficient is {\it not} a universal function of $Q_a$ even in the
limit of ${\tilde L\rightarrow\infty}$ and $\eta\rightarrow 0$,
since the linear density profile persists only in half-space. In
Fig.~2, we show our exact results for several values of
$\zeta=k_0\Lambda$ as functions of the incident angle $\theta$ and
the parameter $Q_a$. As $\zeta$ increases, the curves converge to
the universal shape in Fig.~1.

\begin{figure}
\includegraphics[width=.4\textwidth]{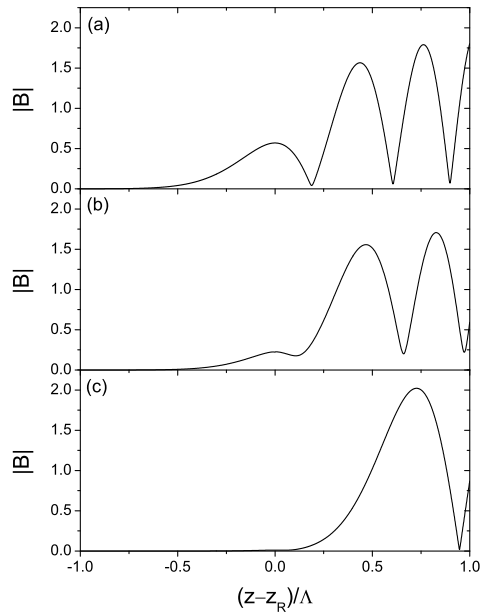}
\caption{Absolute value of the magnetic field, $\vert B\vert$, as a
function of $\left(z-z_R\right)/\Lambda$ for Model B, where $z_R$
($\equiv L-\Lambda$) is the coordinate of the mode conversion point.
$\zeta$ is equal to 12.5. (a) $\theta=5^\circ$ (b)
$\theta=\sin^{-1}0.4\approx 23^\circ$ (c) $\theta=45^\circ$}
\end{figure}

We have also calculated the spatial dependence of the magnetic field
amplitude $B(z)$ using the invariant imbedding equation
(\ref{eq:fd}). In Fig.~3, we plot the absolute value of the magnetic
field amplitude, $\vert B\vert$, for $\zeta=12.5$ and several values
of the incident angle $\theta$ as a function of
$\left(z-z_R\right)/\Lambda$ where $z_R$ ($\equiv L-\Lambda$) is the
coordinate of the mode conversion point. After a suitable rescaling
of the variables, our Fig.~3(b) agrees quite well with Fig.~2(f) in
Ref.~24.

\section{Results on Model C: Parabolic density profile in entire space}
\label{sec:para}

The mode conversion in a warm, unmagnetized plasma with a parabolic
density profile in entire space has been studied analytically by
Hinkel-Lipsker {\it et al.}.\cite{hink0,hink2,hink3} If we restrict
our attention to the cold plasma case with zero temperature, our
Model C is identical to the models studied in Refs.~26$-$28. In
order to obtain the mode conversion coefficient, we solve the
invariant imbedding equations for $r$ and $t$, Eq.~(\ref{eq:rtp0}),
with the initial conditions $r(0)=0$ and $t(0)=1$. In the limit
where ${\tilde L}\rightarrow\infty$ and $\eta\rightarrow 0$, we find
numerically that the mode conversion coefficient ${\mathcal A}$
($=1-\vert r\vert^2-\vert t\vert^2$) is a {\it universal} function
of the parameters $Q_c\equiv
k_0\Lambda\sin^2\theta/(1+\delta)^{1/2}$ and $\Delta^2\equiv
k_0\Lambda\delta/(1+\delta)^{1/2}$. That this has to be the case can
be seen by rewriting the wave equation using the variables
introduced in Ref.~28, ${\overline z}\equiv
{k_0}^{1/2}(1+\delta)^{1/4} z/\Lambda^{1/2}$, ${\overline L}\equiv
{k_0}^{1/2}(1+\delta)^{1/4} L/\Lambda^{1/2}$, $\eta_c\equiv
k_0\Lambda\eta/(1+\delta)^{1/2}$:
\begin{eqnarray}
&&{{{d^2} B}\over{d{\overline z}^2}}-\frac{2\left({\overline
z}-{\overline L}/2\right)}{\left({\overline z}-{\overline
L}/2\right)^2-\Delta^2+i\eta_c}{{dB}\over{d{\overline
z}}}\nonumber\\
&&~~~~+\left[\left({\overline z}-{\overline
L}/2\right)^2-\Delta^2+i\eta_c-Q_c\right]B=0. \label{eq:pc}
\end{eqnarray}

\begin{figure}
\includegraphics[width=.4\textwidth]{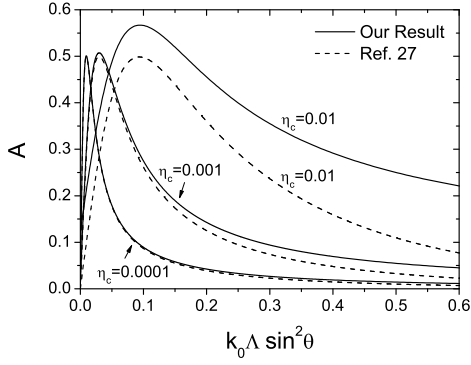}
\caption{Absorption coefficient $\mathcal A$ for Model C as a
function of $Q_c=k_0\Lambda\sin^2\theta$, for several values of the
damping parameter $\eta_c$. The wave frequency is fixed at the peak
plasma frequency, therefore $\delta$ is equal to zero. The value of
$\tilde L$ used in the calculation is 400. Our exact result is
compared with the analytical result of Ref.~27 represented by dashed
lines.}
\end{figure}

\begin{figure}
\includegraphics[width=.4\textwidth]{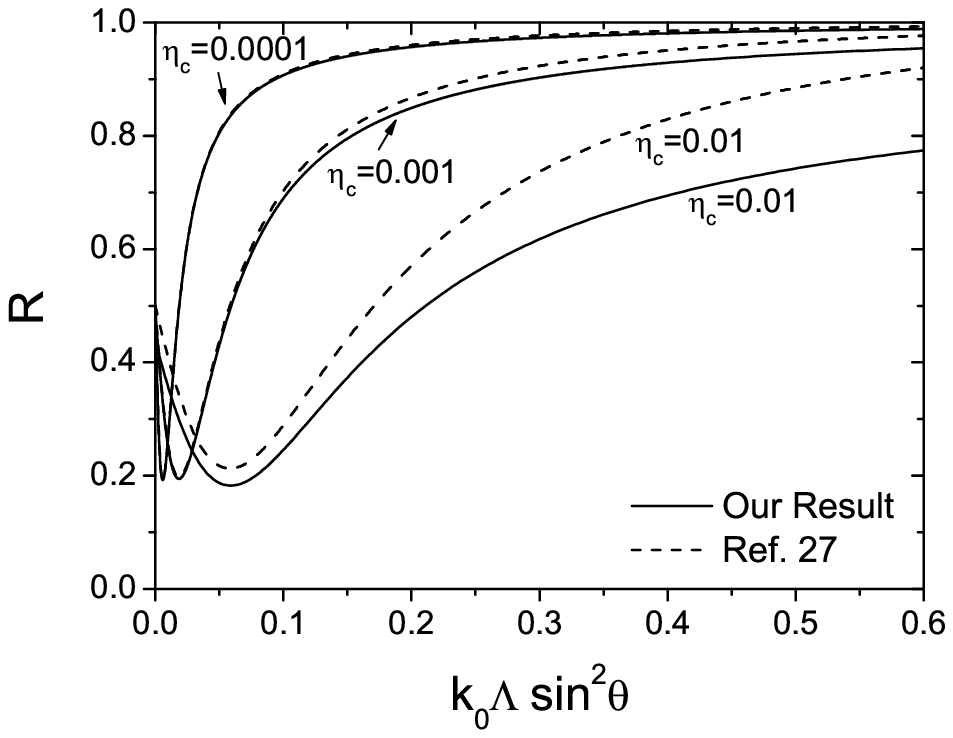}
\caption{Reflectivity $\mathcal R$ for Model C as a function of
$Q_c$, for several values of $\eta_c$. $\delta$ is zero and $\tilde
L$ is 400. Our result is compared with that of Ref.~27 represented
by dashed lines.}
\end{figure}

\begin{figure}
\includegraphics[width=.4\textwidth]{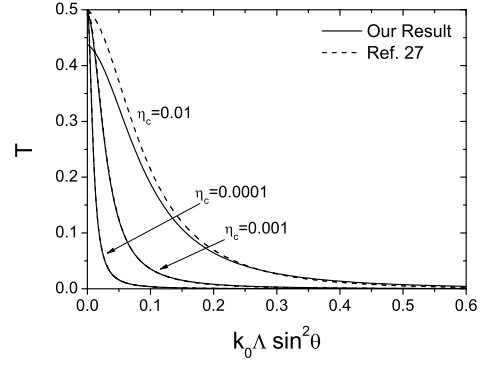}
\caption{Transmissivity $\mathcal T$ for Model C as a function of
$Q_c$, for several values of $\eta_c$. $\delta$ is zero and $\tilde
L$ is 400. Our result is compared with that of Ref.~27 represented
by dashed lines.}
\end{figure}

When the wave frequency is equal to the local plasma frequency at
the top of the parabolic density profile, corresponding to
$\delta=0$, the quantity $\mathcal A$ vanishes unless $\eta_c>0$ and
there is no true mode conversion. In this case, we will call
$\mathcal A$, which measures the absorption due to collisional
damping, as the absorption coefficient instead of the mode
conversion coefficient.

In Figs.~4, 5 and 6, we show the absorption coefficient, the
reflectivity and the transmissivity when $\delta=0$ as a function of
the parameter $Q_c$ for several values of $\eta_c$. The scaled
thickness of the plasma $\tilde L$ used in the calculation is equal
to 400. Our numerically exact results are compared with those in
Ref.~27. The agreement is quite good when $\eta_c=0.0001$, but
becomes worse as $\eta_c$ increases. In fact, if we do the
calculation for larger $\tilde L$ values, we find our absorption
coefficient gets bigger and the discrepancy between our result and
that of Ref.~27 increases gradually. This is reasonable because the
damping occurs in the entire region of thickness $\tilde L$ in our
calculation, whereas several approximations limiting the effects of
damping to a small region surrounding the peak point have been made
in Ref.~27. We have calculated the absorption coefficient for the
cases where the damping parameter $\eta_c$ is nonzero only in a
narrow region of small width surrounding the point $z=L/2$. The
discrepancy between the result of this calculation and that of
Ref.~27 is much smaller than in Fig.~4, but there is no unique way
of choosing a particular width.

\begin{figure}
\includegraphics[width=.4\textwidth]{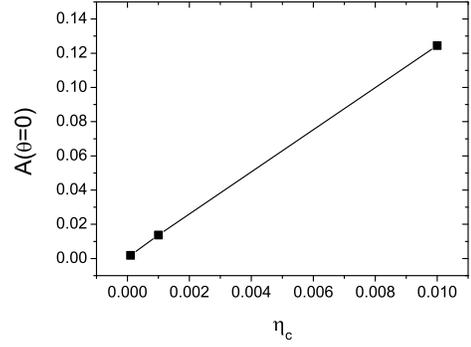}
\caption{Absorption coefficient when the angle of incidence is zero,
$\mathcal A(\theta=0)$, for Model C as a function of the damping
parameter $\eta_c$. $\delta$ is zero and $\tilde L$ is 400.}
\end{figure}

In Fig.~7, we plot the value of the absorption coefficient when the
angle of incidence is zero as a function of the damping parameter.
$\delta$ is zero and $\tilde L$ is 400. We find an almost linear
increase of $\mathcal A(\theta=0)$, whereas it remains zero in
Ref.~27.

\begin{figure}
\includegraphics[width=.4\textwidth]{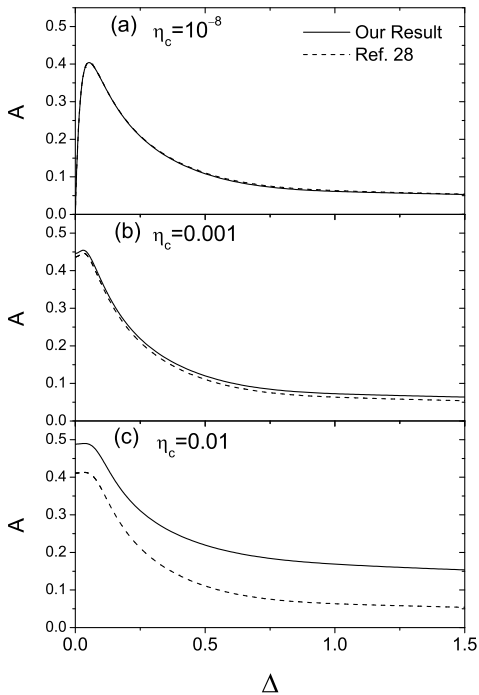}
\caption{Absorption coefficient $\mathcal A$ for Model C as a
function of the parameter $\Delta$ for several values of $\eta_c$.
$Q_c$ is 0.05 and $\tilde L$ is 400. Our result is compared with
that of Ref.~28 represented by dashed lines. (a) $\eta_c=10^{-8}$
(b) $\eta_c=0.001$ (c) $\eta_c=0.01$}
\end{figure}

In Fig.~8, we show the absorption coefficient as a function of the
parameter $\Delta$ for several values of the damping parameter.
$\tilde L$ is 400 and $Q_c$ is 0.05. Our exact result is compared
with the analytical result of Ref.~28. The agreement is quite good
when $\eta_c=10^{-8}$, but becomes worse as $\eta_c$ increases for
the same reason as in Figs.~4, 5 and 6. If $\Delta$ is nonzero, the
wave frequency is shifted from the peak plasma frequency and the
mode conversion occurs at two points,
$z=L/2\pm\left[\delta/\left(\delta+1\right)\right]^{1/2}\Lambda$,
where $\delta$ is related to $Q_c$ and $\Delta$ by
$\delta=\left(\Delta^2\sin^2\theta\right)/Q_c$. For a finite
$\eta_c$, the quantity $\mathcal A$ includes the absorption due to
both collisional damping and mode conversion. As $\eta_c$ approaches
zero, $\mathcal A$ becomes the genuine mode conversion coefficient.
The curve in Fig.~8(a) is a well-converged one which shows only the
absorption due to mode conversion. The differences between Fig.~8(a)
and Figs.~8(b) and 8(c) represent the absorption due to collisional
damping.

\begin{figure}
\includegraphics[width=.4\textwidth]{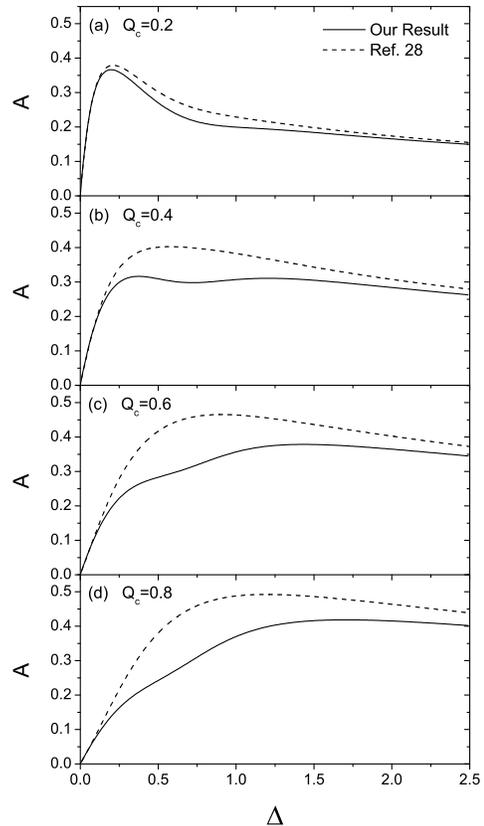}
\caption{Mode conversion coefficient $\mathcal A$ for Model C as a
function of the parameter $\Delta$ for several values of $Q_c$.
$\eta_c$ is $10^{-8}$ and $\tilde L$ is 400. Our result is compared
with that of Ref.~28 represented by dashed lines. (a) $Q_c=0.2$ (b)
$Q_c=0.4$ (c) $Q_c=0.6$ (d) $Q_c=0.8$}
\end{figure}

In Fig.~9, we plot the mode conversion coefficient, which is the
absorption coefficient in the $\eta_c\rightarrow 0$ limit, as a
function of $\Delta$ for several values of $Q_c$. The values of
$\tilde L$ and $\eta_c$ used in the calculation are 400 and
$10^{-8}$ respectively. We have verified that these results are
well-converged in the sense that the results of the calculation for
$\tilde L$ values bigger than 400 and $\eta_c$ values smaller than
$10^{-8}$ are indistinguishable from those shown in Fig.~9. We
believe this is the first exact calculation of the mode conversion
coefficient in the parabolic case. Our exact result is compared with
the analytical formulas of Ref.~28. Unlike the $Q_c=0.05$ case shown
in Fig.~8(a), the discrepancy between our result and that of Ref.~28
is sizable. The shape of the curves is not very simple. For
instance, we observe a double-peaked structure in Fig.~9(b).

\section{Conclusion}
\label{sec:conc}

In this paper, we have presented a new version of the invariant
imbedding theory of the wave propagation in stratified media and
applied it to the mode conversion phenomena of $p$ waves in cold,
unmagnetized plasmas. We have obtained the mode conversion
coefficient and the field distribution exactly for linear and
parabolic plasma density profiles for the first time. Using a recent
generalization of the invariant imbedding theory for the propagation
of coupled waves in stratified media, we can apply our method to
more general situations in a straightforward manner.\cite{kim3} In
forthcoming papers, we will present the result of our study on the
mode conversion in magnetized plasmas and the temperature effects on
both unmagnetized and magnetized plasmas.

\acknowledgments
This work has been supported by the Korea Science
and Engineering Foundation through grant number
R14-2002-062-01000-0.

\end{document}